\documentclass[showpacs,preprintnumbers]{revtex4}
\usepackage{amsmath}
\usepackage{graphicx}
\usepackage{epsf}
\usepackage{psfrag}
\usepackage{epsfig}
\usepackage{graphics}

\begin{document}

\title{Weak Values and the quantum phase space}
\date{\today}
\author{A. C. Lobo}
\email{lobo@iceb.ufop.br}
\author{C. A. Ribeiro}
\email{clyffe@fisica.ufmg.br}
\affiliation{Federal University of Ouro Preto - Physics Department - ICEB \\
Ouro Preto MG - Brazil}

\begin{abstract}
We address the issue of how to properly treat, and in a more general
setting, the concept of a weak value of a weak measurement in quantum
mechanics. We show that for this purpose, one must take in account the
effects of the measuring process on the entire phase space of the measuring
system. By using coherent states, we go a step further than Jozsa in a
recent paper, and we present an example where the result of the measurement
is symmetrical in the position and momentum observables and seems to be much
better suited for quantum optical implementation.
\end{abstract}

\maketitle

\section{Introduction}

The concept of a \textit{weak value} of a quantum mechanical system
was introduced in 1988 by Aharonov, Albert and Vaidman \cite{[1],
aharonov1990}. It was built on a \textit{time symmetrical model }for quantum
mechanics previously introduced by Aharonov, Bergmann and Lebowitz in 1964 %
\cite{[2]}. In this model, non-local time boundary conditions are used,
since the description of the state of a physical system between two quantum
mechanical measurements is made by pre and post-selection of the states. The
authors developed the so called \textit{ABL Rule} for the transition
probabilities within this scenario, so this is why it is also known as the 
\textit{two state formalism} for quantum mechanics \cite{[3]}. The \textit{%
weak value} of an observable can be considered as a generalization of the
usual expectation value of a quantum observable, but differently from this,
it takes values in the \textit{complex plane} in general. Recently, Jozsa
presented a deeper understanding of the physical meaning of the \textit{real}
and \textit{imaginary} part of the complex value \cite{[4]}. In this letter,
we review some of Jozsa's ideas and suggest further progress in the
comprehension of the weak value through a more general analysis of the
effect of the weak measuring process on the quantum phase plane of the
measuring system. Every measurement (whether a common or a weak measurement)
can be understood as an interaction of the system being measured with the
measuring system (the ``measuring device'') described by the von-Neumann
model \cite{[5]}. In the limit of an infinitesimally small coupling of the
measuring system to the system to be measured, the first system will be
accordingly ``infinitesimally perturbed'', but this small effect can be
revealed by taking an average over the measurements of a very large ensemble
of identically prepared systems with the same pair of pre and post selected
states. We propose a more general analysis of the physical meaning of a weak
value than those found in current literature (to the extent of our
knowledge) through a quantum
phase space description of the measuring system. In sections II and III, we briefly
review the concept of the ideal von-Neumann measurement model and the
quantum phase space formalism through coherent states. In section IV, we
review the weak value concept within the quantum phase space formalism
and state the main result of this work. In section V, we address some
concluding remarks and suggestions for further research.

\section{The von-Neumann model for an ideal measurement}

\label{von-neumann}

Let $W=W_{S}\otimes W_{M}$ be the state vector space of the system formed by
the subsystem $W_{S}$ and the \textit{measuring }subsystem $W_{M}$. Assume
further, that we are interested in measuring a discrete quantum variable of $%
W_{M}$ defined by the observable $\widehat{O}=\sum_{i}|o_{i}\rangle o_{i}\langle
o^{i}|$ and that the measuring
subsystem, for simplification purposes, will be considered as a
structureless (no spin or internal variables) quantum mechanical particle in
one dimension. Thus, we can choose as a basis for the vector state space $%
W_{M}$ either of the usual eigenstates of position or momentum $%
\{|q(x)\rangle \}$ or $\{|p(x)\rangle \}$. It is important to note here that
we use a slightly different notation than usual (for reasons that will soon
become evident) in the sense that we distinguish between the ``type'' of the
eigenvector ($q$ or $p$) from the actual $x$ eigenvalue \cite{[8]}. For
instance, we write: 
\begin{equation}
\widehat{Q}|q(x)\rangle =x|q(x)\rangle \qquad \text{and}\qquad \widehat{P}%
|p(x)\rangle =x|p(x)\rangle ,
\end{equation}%
(instead of $\widehat{Q}|q\rangle =q|q\rangle $ and $\widehat{P}|p\rangle
=p|p\rangle $ as commonly written) where $\widehat{Q}$ and $\widehat{P}$ are
the position and momentum observables subject to the well known Heisenberg
relation: $[\widehat{Q},\widehat{P}]=i\widehat{I}$ (hereinafter, $\hbar =1$
units will be used). With this non-standard notation, the completeness
relation and the overlapping between these bases can be written as: 
\begin{equation}
\int\limits_{-\infty }^{+\infty }|q(x)\rangle \langle
q(x)|dx=\int\limits_{-\infty }^{+\infty }|p(x)\rangle \langle p(x)|dx=%
\widehat{I}\quad \text{and}\quad \langle q(x)|p(x\prime )\rangle =\dfrac{%
e^{ixx\prime }}{\sqrt{2\pi }}.
\end{equation}

An\textit{\ ideal} von-Neumann measurement can be defined as an \textit{%
instantaneous} interaction between the two subsystems as modeled by the
following Dirac delta-like time-pulse hamiltonian operator at time $t_{0}$: 
\begin{equation}
\widehat{H}_{int}(t)=\lambda \delta (t-t_{0})\widehat{O}\otimes \widehat{P},
\end{equation}%
where $\lambda $ is a parameter that represents the intensity of the
interaction. This ideal situation models a setup where we are supposing that
the time of interaction is very small compared to the time evolution given
by the free Hamiltonians of both subsystems.

Let the initial state of the total system be given by the following
uncorrelated product state: $|\psi _{(i)}\rangle =|\alpha \rangle \otimes
|\varphi _{(i)}\rangle $ and the final state given by $|\psi _{(f)}\rangle =%
\widehat{U}(t_{A},t_{B})|\psi _{(i)}\rangle \quad (t_{A}<t_{0}<t_{B})$, where
the total unitary evolution operator is%
\begin{equation}
\widehat{U}(t_{A},t_{B})=e^{-i\int_{t_{A}}^{t_{B}}\widehat{H}%
_{int}(t)dt}=e^{-i\lambda \widehat{O}\otimes \widehat{P}},
\end{equation}%
such that 
\begin{equation}
(\widehat{I}\otimes \langle q(x)|)|\psi _{(f)}\rangle = \sum_{j}|o_{j}\rangle \otimes
\langle q(x)|\widehat{V}_{o_{j}}^{\dagger }|\varphi _{(i)}\rangle \alpha ^{j},
\end{equation}%
where $|\alpha \rangle =\sum_{j}|o_{j}\rangle \langle o^{j}|\alpha \rangle
=\sum_{j}|o_{j}\rangle \alpha ^{j}$ and $\widehat{V}_{\xi }$ is the one-parameter
family of unitary operators in $W_{M}$\ that represents the \textit{Abelian
group of translations} in the position basis ($x\in \Re $): 
\begin{equation}
\widehat{V}_{\xi }|q(x)\rangle =|q(x-\xi )\rangle 
\end{equation}%
A correlation in the final state of the total system is then established
between the variable to be measured $o_{j}$ with the continuous position
variable of the measuring particle: 
\begin{equation}
(\widehat{I}\otimes \langle q(x)|)|\psi _{(f)}\rangle =\sum_{j}|o_{j}\rangle \alpha
^{j}\varphi _{(i)}(x-\lambda o_{j}),
\end{equation}%
where $\varphi _{(i)}(x)=\langle q(x)|\varphi _{(i)}\rangle $ is the
wave-function in the position basis of the measuring system (the 1-D
particle) in its initial state. This step of the von-Neumann measurement
prescription is called the \textit{pre-measurement} of the system. The true
measurement happens effectively when an observation of the measuring system
(which is considered to be ``classical''\hspace{1mm}in some sense) is
carried out. If this is the case, one obtains an observed value $x-\lambda
o_{j}$ of the position of the classical particle with a probability $%
P_{j}=|\alpha _{j}|^{2}$. In the Copenhagen interpretation of quantum
physics, the existence of a ``classical description'' of reality is \textit{%
aprioristically} necessary for the description of quantum reality. This kind
of pragmatic and extreme positivistic position, advocated by many physicists
(Bohr and Landau just to mention some of the most prominent) has always been
the center of a heated debate from the very beginning of quantum theory \cite%
{Jammer}. In the last twenty years, a program that tries to offer a solution
to this problem through a full quantum description of the measurement
process making use of the concept of \textit{decoherence} (the inevitable
entangling between the measuring system and the environment) has emerged
resulting in a number of important achievements in various aspects of the
theory. In this work, we are not concerned with these difficult and
foundational aspects of the measurement process. But we should mention that
a full and completely agreed upon resolution of the \textit{measurement
problem} in quantum mechanics does not seem to have been proposed yet \cite%
{[7]}.

\section{The quantum phase space}

By a \textit{quantum symplectic transform}, we mean a unitary transformation
in $W_{M}$ that implements a representation of the group of area preserving
linear transformations of the classical phase plane. For instance, the usual
Fourier transform operator $\widehat{F}$ represents a $\pi /2$ rotation of
the phase plane. In fact, given a coherent state $\mid p,q\rangle $
representing a point in phase plane, one can show that $\widehat{F}\mid
p,q\rangle =\mid q,-p\rangle $ \cite{[8]}. In 1980, Namias developed the
concept of a \textit{fractionary Fourier transform }which has been used
since then in various applications in optics, signal processing and quantum
mechanics \cite{[9]}. This operator is nothing else but an arbitrary
Euclidean rotation in the phase plane. By \textit{Euclidean}, we mean a
linear transformation that \textit{preserves} the usual metric in $\Re ^{2}$
with positive determinant, that is, the one-parameter Abelian group $SO(2)$.
This, of course, does not exhaust all area preserving linear transformations
of the plane which is the non-Abelian $SL(2,\Re )$ group. We can define the
Fourier transform operator as 
\begin{equation}
\widehat{F}=\int\limits_{-\infty }^{+\infty }dx\mid p(x)\rangle \langle
q(x)|.
\end{equation}%
(Note that it would be impossible to define the Fourier operator in such a
clean and direct way with the usual $\mid q\rangle $ and $\mid p\rangle $
notation for the position and momentum eigenstates). The squared Fourier
operator $\widehat{F}^{2}$ is the space inversion operator and it can be
shown that $\widehat{F}^{3}=\widehat{F}^{\dagger }$ and $\widehat{F}^{4}=%
\widehat{I}$, so it can be clearly seen that the eigenvalues of $\widehat{F}$
are the fourth roots of unity. In fact, it is well known that 
\begin{equation}
\widehat{F}\mid n\rangle =(i)^{n}\mid n\rangle ,
\end{equation}%
where $\{\mid n\rangle \},$ $(n=0,1,2\ldots )$ is the complete set of
eigenkets of the \textit{number operator} $\widehat{N}=\widehat{a}^{t}%
\widehat{a}$, which is in itself, the generator of rotations in the phase
plane \cite{[8]}. Probably the best way to visualize this is through the
identification of the phase plane with the complex plane via the standard
complex-valued coherent states defined by the following change of variables: 
\begin{equation}
z=\frac{1}{\sqrt{2}}(q+ip)
\end{equation}%
and defined as%
\begin{equation}
\mid z\rangle =\widehat{D}[z]\mid 0\rangle ,
\end{equation}%
with the displacement $\widehat{D}[z]$ operator given by 
\begin{equation}
\widehat{D}[z]=e^{(z\widehat{a}^{t}-\overset{\_}{z}\widehat{a})}.
\end{equation}%
It is not difficult then to show that indeed 
\begin{equation}
e^{i\theta \widehat{N}}\mid z\rangle =\mid e^{i\theta }z\rangle .
\end{equation}%
so that, 
\begin{equation}
\widehat{F}_{\theta }=e^{i\theta \widehat{N}}
\end{equation}%
is in fact the Namias fractional Fourier operator and the usual Fourier
operator is the special case with $\theta =\pi /2$. We can rewrite $\widehat{%
N}$ in terms of the position and momentum operators since $\widehat{a}=\frac{%
1}{\sqrt{2}}(\widehat{Q}+i\widehat{P})$, so the number operator may be
expressed as 
\begin{equation}
\widehat{N}=\widehat{H}_{0}-\frac{1}{2}\widehat{I}=\frac{1}{2}(\widehat{Q}%
^{2}+\widehat{P}^{2})-\frac{1}{2}\widehat{I},
\end{equation}%
where $\widehat{H}_{0}$\ is the Hamiltonian of a unit mass and unit
frequency simple harmonic oscillator. Since the generator of rotations is 
\textit{quadratic} in the canonical observables $\widehat{Q}$ and $\widehat{P%
}$, one may try to write down all possible quadratic operators in these
variables: $\widehat{Q}^{2}$, $\widehat{P}^{2}$, $\widehat{Q}\widehat{P}$
and $\widehat{P}\widehat{Q}$, but the last two are obviously non-Hermitian
so we could change them to the following (Hermitian) linear combinations: $%
\widehat{Q}\widehat{P}+\widehat{P}\widehat{Q}$ and $i(\widehat{Q}\widehat{P}-%
\widehat{P}\widehat{Q})$. The last one is proportional to the identity
operator because of the Heisenberg commutation relation, so this leaves us
with \textit{three} linear independent operators that we choose as 
\begin{equation}
\widehat{H}_{0}=\frac{1}{2}(\widehat{Q}^{2}+\widehat{P}^{2})=\widehat{N}+%
\frac{1}{2}\widehat{I}=\widehat{a}^{t}\widehat{a}+\frac{1}{2}\widehat{I},
\end{equation}%
\begin{equation}
\widehat{g}=\frac{1}{2}(\widehat{Q}\widehat{P}+\widehat{P}\widehat{Q})=\frac{%
i}{2}[(\widehat{a}^{t})^{2}-\widehat{a}^{2}],
\end{equation}%
\begin{equation}
\widehat{k}=\frac{1}{2}(\widehat{Q}^{2}-\widehat{P}^{2})=\frac{1}{2}[(%
\widehat{a}^{t})^{2}+\widehat{a}^{2}].
\end{equation}%
These three generators implement in $W_{M}$, the algebra $sl(2,\Re )$ of $%
SL(2,\Re )$. The $\widehat{g}$ operator is nothing but the squeezing
generator from quantum optics \cite{[12]}. The $\widehat{k}$ operator
generates \textit{hyperbolic rotations, }that is, linear transformations of
the plane that preserve an \textit{indefinite metric}. It takes the
hyperbola $x^{2}-y^{2}=1$ into itself in an analogous way that the Euclidean
rotation takes the circle $x^{2}+y^{2}=1$ into itself. $SL(2,\Re )$ is the
Lie Group of all area preserving linear transformations of the plane, so we
can identify it with the $2\times 2$\ real matrices with unit determinant.
Since $\det e^{X}=e^{tr(X)}$, we can also identify its algebra $sl(2,\Re )$
with all $2\times 2$\ real matrices with null trace. Thus, it is natural to
make the following choice for a basis in this algebra: 
\begin{equation*}
\mathbf{X_{1}}=\left( 
\begin{array}{cc}
0 & 1 \\ 
1 & 0%
\end{array}%
\right) =\widehat{\sigma }_{1}\qquad \mathbf{X_{2}}=\left( 
\begin{array}{cc}
0 & -1 \\ 
1 & 0%
\end{array}%
\right) =-i\widehat{\sigma }_{2}\qquad \mathbf{X_{3}}=\left( 
\begin{array}{cc}
1 & 0 \\ 
0 & -1%
\end{array}%
\right) =\widehat{\sigma }_{3},
\end{equation*}%
where we have written (for practical purposes) the elements of the algebra
in terms of the well-known \textit{Pauli matrices}. This is very adequate
because physicists are familiar with the fact that the Pauli matrices $\{%
\frac{1}{2}\widehat{\sigma }_{j}\}$ form a two-dimensional representation of
the angular momentum algebra:%
\begin{equation*}
\lbrack \widehat{\sigma }_{i},\widehat{\sigma }_{j}]=2i\widehat{\sigma }%
_{k}\epsilon _{ij}^{k}.
\end{equation*}%
We can make use of these commutation relations to completely characterize
the $sl(2,\Re )$ algebra. In fact, the mapping described by the table below
relates these algebra elements directly to the algebra of their
representation carried on $W_{M}$: 
\begin{equation}
\begin{tabular}{llllll}
\textbf{generators of }$sl(2,\Re )$ &  &  &  &  & \textbf{generators of the
representation} \\ 
$X_{1}\equiv \widehat{\sigma }_{1}$ &  &  &  &  & $-i\widehat{k}$ \\ 
$X_{2}\equiv -i\widehat{\sigma }_{2}$ &  &  &  &  & $-i\widehat{H}_{0}$ \\ 
$X_{3}\equiv \widehat{\sigma }_{3}$ &  &  &  &  & $-i\widehat{g}$%
\end{tabular}
\label{19}
\end{equation}%
With a bit of work, it is not difficult to convince oneself that these
mapped elements indeed obey identical commutation relations.

\section{Weak Values}

The \textit{weak value} of a quantum system introduced by Aharonov,
Albert and Vaidman (A.A.V.) based on the two-state formalism for quantum
mechanics generalizes the concept of an expectation value for a given
observable. Let the \textit{initial state} of the product space $%
W=W_{S}\otimes W_{M}$ be $|\psi _{(i)}\rangle =|\alpha \rangle \otimes
|\varphi _{(i)}\rangle $ and a ``\textit{weak Hamiltonian}''given as: 
\begin{equation}
\widehat{H}_{int}^{(w)}(t)=\epsilon \delta (t-t_{0})\widehat{O}\otimes 
\widehat{P},\qquad (\epsilon \rightarrow 0).  \label{20}
\end{equation}%
The \textit{final state} will then be: $|\varphi _{(f)}\rangle =(\langle \beta
|\otimes \widehat{I}).\widehat{U}^{(w)}(t_{i},t_{f}).(|\alpha \rangle
\otimes |\varphi _{(i)}\rangle )$, where $|\alpha \rangle $ and $|\beta
\rangle $ are respectively the \textit{pre and post-selected} states of the
system to be measured $W_{M}$ and $\widehat{U}^{(w)}(t_{i},t_{f})\simeq 
\widehat{I}-i\epsilon \widehat{O}\otimes \widehat{P}$ is the \textit{time
evolution} operator for the weak interaction. In this way, we can compute to
first order in $\epsilon $: 
\begin{equation}
|\varphi _{(f)}\rangle \simeq \langle \beta |\alpha \rangle (1-i\epsilon O_{w}%
\widehat{P})|\varphi _{(i)}\rangle \qquad \text{with}\qquad O_{w}=\dfrac{%
\langle \beta |\widehat{O}|\alpha \rangle }{\langle \beta |\alpha \rangle }.
\end{equation}%
Note that the weak value $O_{w}$ of the observable $\widehat{O}$ is, in
general, an arbitrary \textit{complex number}. Note also that, though $%
|\varphi _{(i)}\rangle $ is a normalized state, the $|\varphi _{(f)}\rangle $
state vector in general, \textit{is not} normalized. In the original
formulation of A.A.V, the momentum $\widehat{P}$ acts upon the measuring
system, implementing a \textit{small} \textit{translation} of the initial
wave function in the position basis, but which can be measured from the mean
value of the results of a large series of identical experiments. That is,
the expectation value of the position operator $\widehat{Q}$ over a large
ensemble with the same pre and post selected states. Jozsa recently proposed
a more general procedure by taking an arbitrary operator $\widehat{M}$ in
the place of $\widehat{Q}$ as the observable of $W_{M}$ to be measured. In
this case, the usual expectation values of $\widehat{M}$ in the initial and
final states $|\varphi _{(i)}\rangle $ and $|\varphi _{(f)}\rangle $ are
respectively: 
\begin{equation}
\langle \widehat{M}\rangle _{(i)}=\langle \varphi _{(i)}|\widehat{M}|\varphi
_{(i)}\rangle \qquad \text{and}\qquad \langle \widehat{M}\rangle _{(f)}=\dfrac{%
\langle \varphi _{(f)}|\widehat{M}|\varphi _{(f)}\rangle }{\langle \varphi
_{(f)}|\varphi _{(f)}\rangle }.
\end{equation}%
Jozsa has shown that the difference between these expectation values to
first order in $\epsilon $ is given by \cite{[4]}: (the shift of $\langle 
\widehat{M}\rangle $ is defined as $\Delta \widehat{M}=\langle \widehat{M}%
\rangle _{(f)}-\langle \widehat{M}\rangle _{(i)}$) 
\begin{equation}
\Delta \widehat{M}=\epsilon \lbrack (Im(O_{w}))(\langle \varphi _{(i)}|\{%
\widehat{M},\widehat{P}\}|\varphi _{(i)}\rangle -2\langle \varphi _{(i)}|%
\widehat{P}|\varphi _{(i)}\rangle \langle \varphi _{(i)}|\widehat{M}|\varphi
_{(i)}\rangle )-i(Re(O_{w}))\langle \varphi _{(i)}|\widehat{[M},\widehat{P}%
]|\varphi _{(i)}\rangle ].  \label{23}
\end{equation}

He also discusses two different examples for $\widehat{M}$: At first, he
chooses $\widehat{M}=\widehat{Q}$, so that (using the $sl(2,\Re )$ algebra
and the Heisenberg commutation relation) he obtains: 
\begin{equation}
\Delta \widehat{Q}=\epsilon \lbrack 2(Im(O_{w}))(\langle \varphi _{(i)}|%
\widehat{g}|\varphi _{(i)}\rangle -\langle \varphi _{(i)}|\widehat{P}|\varphi
_{(i)}\rangle \langle \varphi _{(i)}|\widehat{Q}|\varphi _{(i)}\rangle
)+(Re(O_{w}))].
\end{equation}

By using the Heisenberg picture for time evolution and choosing the most
general Hamiltonian for the measuring system $W_{M}$ and the relations of
table \ref{19}. 
\begin{equation}
\widehat{H}_{M}=\frac{1}{2m}\widehat{P}^{2}+V(\widehat{Q})
\end{equation}%
one obtains: 
\begin{equation}
\dfrac{d\widehat{Q}}{dt}=\dfrac{\widehat{P}}{m}\qquad \text{and}\qquad 
\dfrac{d\widehat{Q}^{2}}{dt}=\frac{2}{m}\widehat{g}.
\end{equation}%

So one arrives at:%
\begin{equation}
\Delta \widehat{Q}=\epsilon \lbrack (Re(O_{w}))+m(Im(O_{w}))\dfrac{d}{dt}%
(\delta _{|\varphi _{(i)}\rangle }^{2}\widehat{Q})],  \label{27}
\end{equation}%
where $\delta _{|\varphi _{(i)}\rangle }^{2}\widehat{A}=\langle \varphi _{(i)}|%
\widehat{A}^{2}|\varphi _{(i)}\rangle -\langle \varphi _{(i)}|\widehat{A}%
|\varphi _{(i)}\rangle ^{2}$\hspace{1mm}is the usual quadratic dispersion or
uncertainty of the $\widehat{A}$ observable in an arbitrary state vector $%
|\varphi _{(i)}\rangle $. Analogously for $\widehat{M}=\widehat{P\text{,}}$
comes: 
\begin{equation}
\Delta \widehat{P}=2\epsilon (Im(O_{w}))\delta _{|\varphi _{(i)}\rangle }^{2}%
\widehat{P}.  \label{28}
\end{equation}

Note that there is a certain \textit{asymmetry} in the results exhibited by
the above equations. This is because of the \textit{asymmetric} choice of
the translation generator $\widehat{P}$ in the interaction Hamiltonian in
equation \ref{20}. Note also that from equations \ref{27} and \ref{28} we
can see that it is impossible to extract the \textit{real }and \textit{%
imaginary} values of $O_{w}$ with the measurement of $\Delta \widehat{Q}$ 
\textit{only}, because both of these numbers are absorbed in a \textit{same} 
\textit{real} number. It is necessary to measure $\Delta \widehat{P}$
(besides knowing the values of $\dfrac{d}{dt}(\delta _{|\varphi _{(i)}\rangle
}^{2}\widehat{Q})$ and $\delta _{|\varphi _{(i)}\rangle }^{2}\widehat{P}$).
There is no reason why one should need to choose $\widehat{P}$ or $\widehat{Q%
}$\ in the weak measurement Hamiltonian. We could choose any of the
symplectic generators making use of the full symmetry of the $SL(2,\Re )$
group. The $\widehat{P}$ and $\widehat{Q}$ operators generate translations
in phase space, but we can implement any area preserving transformation in
the plane by also using observables that are quadratic in the momentum and
position observables. We can also make use of our freedom of choice of an
arbitrary initial state vector $|\varphi _{(i)}\rangle $ and the choice of an
``adequate'' Hamiltonian operator $\widehat{H}_{M}$ of the measuring system.
We propose then a more general approach. Let us take an interaction
Hamiltonian of the following form: 
\begin{equation}
\widehat{H}_{int}^{(w)}(t)=\epsilon \delta (t-t_{0})\widehat{O}\otimes 
\widehat{R}\qquad \text{with}\qquad (\epsilon \rightarrow 0),  \label{78}
\end{equation}%
where $\widehat{R}$ is any element of the algebra $sl(2,\Re )$, so it is the
generator of an arbitrary symplectic transform of the measuring system. In
this way, we can follow Jozsa's path obtaining the generalized $\Delta 
\widehat{M}$ shift in these conditions: 
\begin{equation}
\Delta \widehat{M}=\epsilon \lbrack (Im(O_{w}))(\langle \varphi _{(i)}|\{\widehat{%
M},\widehat{R}\}|\varphi _{(i)}\rangle -2\langle \varphi _{(i)}|\widehat{R}|\varphi
_{(i)}\rangle \langle \varphi _{(i)}|\widehat{M}|\varphi _{(i)}\rangle
)-i(Re(O_{w}))\langle \varphi _{(i)}|\widehat{[M},\widehat{R}]|\varphi _{(i)}\rangle ].
\label{79}
\end{equation}

By choosing $\widehat{M}=\widehat{R}$, we get the analog of equation \ref{28}%
: 
\begin{equation}
\Delta \widehat{R}=2\epsilon (Im(O_{w}))\delta _{|\varphi _{(i)}\rangle }^{2}%
\widehat{R}.  \label{80}
\end{equation}%
For the second observable, we could choose any Hermitian operator that 
\textit{does not commute} with $\widehat{R}$. This is because the main idea
is to choose a ``conjugate'' variable to $\widehat{R}$ in a similar way that
occurs with the ($\widehat{Q}$, $\widehat{P}$) pair. So a first obvious
choice is to pick the number operator $\widehat{N}$ in the place of $%
\widehat{R}$. Since $\widehat{N}$ is the generator of Euclidean rotations in
phase space, the annihilator operator $\widehat{a}$ seems a natural
candidate choice to go along with $\widehat{N}$. Though $\widehat{a}$ is not
Hermitean and as such, not a genuine observable, one may think that it would
be useless as such. But any operator $\widehat{B}$ can be written as a sum
of its hermitean and anti-hermitean components in the following manner: 
\begin{equation}
\widehat{B}=\widehat{C}+i\widehat{D},
\end{equation}%
where $\widehat{C}$ and $\widehat{D}$ are both Hermitean. In this manner we
can define the expectation value of any operator in an arbitrary state
vector $|\psi \rangle $ by \cite{[6]}.

\begin{equation}
\langle \widehat{B}\rangle _{|\psi \rangle }=\langle \widehat{C}\rangle
_{|\psi \rangle }+i\langle \widehat{D}\rangle _{|\psi \rangle }.
\end{equation}

Note also that by the linearity of $\widehat{M}$ in equation \ref{23} we
have 
\begin{equation}  \label{34}
\Delta \widehat{B}=\Delta \widehat{C}+i\Delta \widehat{D}.
\end{equation}

So we choose $(\widehat{N}$, $\widehat{a})$ in this manner as a candidate
for a ``conjugate pair'' of operators as an analog to the pair $(\widehat{P}$%
, $\widehat{Q})$ because $\widehat{F}_{\theta }=e^{i\theta \widehat{N}}$
implements Euclidean rotations in phase plane while the coherent state $%
|z\rangle $ is an eigenket of $\widehat{a}$ in a similar way that the
momentum operator implements translations in the position wave function.
With this choice of $\widehat{M}=\widehat{a}$ it is not difficult to
calculate the shift for the annihilator operator: 
\begin{equation}
\Delta \widehat{a}=\epsilon \lbrack -iO_{w}\langle \varphi _{(i)}|\widehat{a}%
|\varphi _{(i)}\rangle +2Im(O_{w})(\langle \varphi _{(i)}|\widehat{N}\widehat{a}|\varphi
_{(i)}\rangle -\langle \varphi _{(i)}|\widehat{N}|\varphi _{(i)}\rangle \langle \varphi
_{(i)}|\widehat{a}|\varphi _{(i)}\rangle )].  \label{81}
\end{equation}

Unfortunately, the second term in the above equation cannot be identified
with a ``quadratic dispersion'' for $\widehat{a}$ in the same way as Jozsa
does for $\widehat{Q}$ in equation \ref{27}. In most models of weak
measurements, the initial state of the measuring system is chosen to be a
Gaussian state and the weak interaction promotes a small translation of its
peak. But in a realistic quantum optical implementation of the measuring
system, it is reasonable to choose the initial state of the system as a
coherent state $|\varphi _{(i)}\rangle =|z\rangle $. In this case, there is a
dramatic simplification for the shift: 
\begin{equation}
\Delta \widehat{a}=-i\epsilon zO_{w}.  \label{36}
\end{equation}%
The above equation is the main result of this work. Note that equation \ref%
{36} can be re-written as: 
\begin{equation}
\Delta \widehat{a}=\epsilon \left| z\right| \left| O_{w}\right| e^{i(\theta
_{z}+\theta _{w}-\pi /2)},
\end{equation}%
where $z=\left| z\right| e^{i\theta _{z}}$ and $O_{w}=\left| O_{w}\right|
e^{i\theta _{w}}$. If we make a convenient choice for the phase $\theta
_{z}=\pi /2$ and use equation \ref{34}, we arrive at a symmetric pair of
equations for $\Delta \widehat{Q}$ and $\Delta \widehat{P}$: 
\begin{equation}
\Delta \widehat{Q}=\epsilon \sqrt{2}\left| z\right| Re(O_{w})
\end{equation}%
and 
\begin{equation}
\Delta \widehat{P}=\epsilon \sqrt{2}\left| z\right| Im(O_{w}).
\end{equation}%
These equations do not depend on the quadratic dispersion or the time
derivative of the quadratic dispersion of any observable as it happens with
the similar equation developed by Jozsa. An additional attractiveness of the
above equations in comparison to those exhibited by Jozsa is the fact that
one can in principle ``tune'' the size of the $\epsilon \left| z\right| $
term despite how small $\epsilon $ may be by making $\left| z\right| $ large
enough. This may be of practical importance for optical implementation of
weak value since $\left| z\right| $ for a quantized mode of an
electromagnetic field is nothing else but the \textit{mean photon number} in
this mode in the $|z\rangle $ coherent state \cite{[12]}. One may envisage
an experiment with a drastic reduction of the size of the ensemble, maybe
even measuring the weak value with \textit{one single} experiment.

\section{Conclusions}

The main message of our work is to underline the fact that the best way to
understand the meaning of the complex weak value is to seek for
the effects upon the phase space of the measuring system. We have shown an
improvement of this understanding by choosing an experiment where the
observable of the measuring system of the weak interaction Hamiltonian is
chosen to be $\widehat{R}=\widehat{N}$ and the ``variable'' chosen to be
measured is $\widehat{M}=\widehat{a}$. This seems a natural choice in a
quantum optical experimental framework. Instead of looking to a translation
of a Gaussian state as the usual proposals, we look to a ``turn of the
dial'' of our coherent state $|z\rangle $ as a \textit{pointer state}. It
puts on a same basis the \textit{real} and \textit{complex} parts of the so
called weak value in a natural way. We could indeed try to explore the full
potentiality of all symplectic transforms in phase space, but he have chosen
to start with the Euclidean rotation (or fractional Fourier transform)
because of its very vivid and simple geometrical interpretation with the use
of coherent states. We shall address, in a future paper, the full
mathematical structure of all the symplectic group and its implementations %
\cite{[13]}. Though the two state formalism of quantum mechanics appeared in
the sixties decade of the last century, it is fair to say that it remained
quite unnoticed until the concept of weak values was introduced by
Aharonov, Albert and Vaidman in 1988. This novel idea has already shown a
great deal of applications, but in our opinion the most interesting and
important one is related to the phenomenon known as \textit{quantum
counterfactuality}. A widely known example is that of the \textit{%
Elitzur-Vaidman bomb problem}, where a very sensitive photon detector (the
``bomb'') is itself ``detected'' by a photon without having really
interacted with it. The fact that the paths of two distinct open quantum
channels can interfere destructively (in the Feynman sense) allows the
possibility of ``detecting'' a sensor which interrupts one of the channels
in an interaction-free like experiment \cite{[10], [11]}. Problems like
these have been successfully analyzed through the concept of weak
values, because these different paths may be actually ``tested''
without significantly perturbing the involved states. We believe that the
understanding of this kind of approach to the study of quantum
counterfactuality also could be enhanced by a shift to the general analysis
of the effect on the phase space of the measuring systems that we are
proposing. In this sense, it may be useful to consider discrete phase spaces
as well \cite{[8]}. In quantum optical applications, observables that are
quadratic on the creation and annihilation operators are a routine matter,
so we believe that this more general quantum phase space approach opens
space for a gain of flexibility and a welcome increase of theoretical and
experimental options for further investigations.

\section{Acknowledgements}

ACL acknowledges the many useful commentaries of Jos\'{e} C\'{e}sar Lobo
Filho on the correct English grammar and style of the text. We also thank the anonymous referee for many useful comments and in particular for have carefully pointed out the distinction between the concepts of a weak value and a weak measurement.

\end{document}